\documentclass[letterpaper, 10 pt, conference]{ieeeconf}

\IEEEoverridecommandlockouts
\usepackage[utf8]{inputenc}
\usepackage[T1]{fontenc}
\usepackage{cite}
\usepackage{amsmath,amssymb,amsfonts}
\usepackage{algorithmic}
\usepackage{graphicx}
\usepackage{textcomp}
\usepackage{amsmath}
\usepackage{amsfonts}
\usepackage{amssymb}
\usepackage{siunitx}
\usepackage{braket}
\usepackage{verbatim}
\usepackage{hyperref} 

\title{\LARGE \bf Memory-optimised Cubic Splines for High-fidelity Quantum Operations}
\author{
    Jan Ole Ernst,$^{1, 2}$
    Jan Snoeijs,$^{1}$
    Mitchell Peaks,$^{1}$ and
    Jochen Wolf$^{1}$\vspace{0.1cm}
    \\ 
    \small\textit{$^{1}$Riverlane, Cambridge CB2 3BZ, UK} \\
    \small\textit{$^{2}$Clarendon Laboratory, University of Oxford, Parks Road, Oxford OX1 3PU, UK} \\
    \small Email: snoeijs.jan@gmail.com}

\begin{document}

\maketitle
\thispagestyle{empty}
\pagestyle{empty}

\begin{abstract}
    Radio-frequency pulses are widespread for the control of quantum bits and the execution of operations in quantum computers. The ability to tune key pulse parameters such as time-dependent amplitude, phase, and frequency is essential to achieve maximal gate fidelity and mitigate errors. As systems scale, a larger fraction of the control electronic processing will move closer to the qubits, to enhance integration and minimise latency in operations requiring fast feedback. This will constrain the space available in the memory of the control electronics to load time-resolved pulse parameters at high sampling rates. Cubic spline interpolation is a powerful and widespread technique that divides the pulse into segments of cubic polynomials. We show an optimised implementation of this strategy, using a two-stage curve fitting process and additional symmetry operations to load a high-sampling pulse output on an FPGA. This results in a favourable accuracy versus memory footprint trade-off. By simulating single-qubit population transfer and atom transport on a neutral atom device, we show that we can achieve high fidelities with low memory requirements. This is instrumental for scaling up the number of qubits and gate operations in environments where memory is a limited resource. \footnote{This work has been submitted to the IEEE for possible publication. Copyright may be transferred without notice, after which this version may no longer be accessible.}
\end{abstract}


\section{Introduction} 
Quantum computing is a powerful new computing paradigm. Yet to unlock its value for practical problems, devices need to scale significantly beyond the current state of the art. This motivates a study of the classical computational resources required to store and represent radio-frequency signals, which are used ubiquitously in most quantum computing architectures to execute operations on the qubits. Various platforms have shown success, with some of the most mature architectures being superconducting qubits, trapped ions, and neutral atoms. The pulse engineering techniques described in this paper are platform-agnostic but particularly suitable for pulses that are smooth, continuous (in both frequency and amplitude), and on the order of $\mu s$. 

The pulse shapes used to address atomic or ionic qubits, both for the implementation of gates and for moving atoms spatially in dipole traps, tend to be smooth and continuous in both their amplitude and frequency envelopes. This is what is typically required for implementing high-fidelity quantum operations. Gate durations vary dramatically across different experimental platforms. Generally, state-of-the-art single- and multi-qubit gates require pulses on the order of microseconds \cite{saner2023breaking, parallel2019}, with some faster gates, such as Rydberg gates, happening on the order of hundreds of nanoseconds \cite{Jandura2022timeoptimaltwothree, Evered2023}, and some longer microwave gates stretching up to milliseconds \cite{Hensinger2015, Harty2016}. Even longer pulses of up to several milliseconds are required to move neutral atoms that have been randomly loaded into a large static array of dipole traps. This is required to create defect-free topologies for analogue quantum simulation or fault-tolerant quantum computation \cite{Schymik2020, Endres2016, barredo2016}, as well as to move atoms among the traps to achieve local entanglement \cite{Bluvstein2022}. The relevance of the pulse compression techniques introduced here to superconducting qubits or other material platforms is discussed in the outlook and left for future work as we focus specifically on neutral atoms in Sec. \ref{sec:benchmarking}.

Synthesising the pulses at sufficient instantaneous bandwidth (typically hundreds of MHz to 1 GHz) typically requires a few kilobits to tens or hundreds of kilobits for pulses lasting over a microsecond if the modulation data is fully stored in local memories inside the control systems' FPGAs. As pulses are typically engineered and calibrated on a per-channel basis, pulse data storage can require a significant amount of memory in scaled-out quantum computing systems. Additionally, FPGA memory loading time is correlated with its memory size, regardless of the data-transfer mechanism. Compression techniques aim to reduce the memory footprint of pulses with little to no error with respect to the original non-compressed pulse representation and could benefit the integration of large-scale control electronics \cite{cryo_cmos}.

A powerful and well-known technique used to represent smooth, continuous functions is the use of cubic spline interpolation by partitioning the function into segments of cubic polynomials \cite{Schoenberg1946ContributionsTT, Bartels1987, birkhoff1965piecewise}. Here we review the existing cubic spline fitting techniques and show how the position of the segment boundaries plays an important role in the accuracy of the spline interpolation; we introduce some heuristic optimisation methods for the position of the knots and show how pulse symmetry can be used to compress the pulse information. Thereafter, we describe our implementation of the cubic spline decompression on an FPGA and show how this leads to cumulative errors, arising from the truncation of the cubic polynomial coefficients to a fixed-point representation on the hardware. This strongly restricts the possible segment length and achievable compression factors. To rectify this, we introduce a novel quantisation-aware cubic spline fitting technique, with which we can achieve a much-improved pulse representation with a comparatively simpler hardware implementation. Using these curve fitting techniques, we benchmark the pulse compression on a simulator for a neutral atom single-qubit gate, as well as an atom transport sequence, and show that the compression introduces almost no reduction in fidelity and achieves a significantly improved ratio of fidelity to pulse memory. This is important for memory-sparse settings such as cryogenic or compact integrated control systems.

\section{Methods}

\subsection{Floating point fit}
\label{sec:floating_point_fit}
Fitting smooth functions with piece-wise cubic polynomials has been well studied for decades \cite{Schoenberg1946ContributionsTT}. Consider a smooth function $y(t)$ (alternatively, this could also be a number of $(t_i,y_i)$ pairs) defined over a range $[t_i,t_f]$. A cubic spline interpolation with $k+1$ piece-wise polynomials and $k$ knots is formulated as follows:
\begin{equation}
\begin{aligned}
        y(t):= \begin{cases}
            P(t,\vec{p_0}), \text{ for } t_i \leq t \leq \tau_1 \\
            P(t,\vec{p_1}), \text{ for } \tau_2 \leq t \leq \tau_3 \\
            \vdots \\
            P(t,\vec{p_k}), \text{ for } \tau_k \leq t \leq t_f
        \end{cases}
\end{aligned}
\label{eq:cubic_spline_def}
\end{equation}
where $\vec{\tau}=(\tau_1,...,\tau_k)$ is the ordered knot vector and the polynomial $P$ is defined as $P(t, \vec{p})= p^0 + p^1t +p^2 t^2 + p^3 t^3$ where $\vec{p_i}$ is the polynomial coefficient vector. When performing the fitting, it is possible to impose not only continuity of first- but also second- or higher-order derivatives at the knot positions $\tau_i$ \footnote{When moving to an FPGA, time will be quantised in terms of the sampling frequency of the output, whereas here it is treated as continuous.}. Another important choice for the fitting routine is the error function. Arguably, the most common error function is the least squares error function $\epsilon = \sum_i(P(t)-y_i)^2$ for desired data points $y_i$ and points generated by the cubic polynomial $P(t)$, but one can also consider an absolute linear error \cite{deBoor1973_splines}. In what follows we will adopt the least squares error criterion since it typically leads to more accurate fits for sparsely sampled higher-order polynomials and, more importantly, this allows the error to be treated as a continuous and differentiable quantity which is instrumental for gradient-based minimisation algorithms for curve fitting. In general, using a larger number of knots and shorter segments improves the accuracy of the fit, but it should also be noted that the memory load of the pulse representation is linear in the number of knots (a brief discussion of the optimisation of the knot positions is given in App. \ref{sec:app_knots}).

In our experience, typical pulse shapes used for implementing quantum gates and other quantum operations such as neutral atom shuttling often exhibit waveforms that have some symmetry in their amplitude or phase envelope. In particular, pulse symmetry around the centre of the amplitude envelope is common for many pulses such as Gaussians (cf. \cite{Bluvstein2023}), Sinusoids (cf. \cite{mixed_species_entangling}), or Blackman ramps (cf. \cite{Ebadi2021}) for atom shuttling waveforms. We can utilise such symmetries to further enhance pulse compression. By dividing the pulse envelope (this can be either an amplitude or phase envelope, but amplitude is more common) into an even number of $N_s$ segments with the $N_s/2$th segment ending exactly at the symmetry centre, one can reduce the required number of stored segments by one half \footnote{For every symmetry that is exploited.}. For example, a symmetric Blackman envelope is divided into 8 segments, but only the first 4 of those need to be stored, as every point past the symmetry centre can be calculated, by a reflection. This symmetry-enhanced compression method can be implemented on the FPGA by effectively replaying the first half of the stored pulse by inverting the time direction, as described in Sec. \ref{sec:fpga}.

\subsection{Implementation on FPGA}
\label{sec:fpga}

\begin{figure}[t!]
\includegraphics[width=0.99\linewidth]{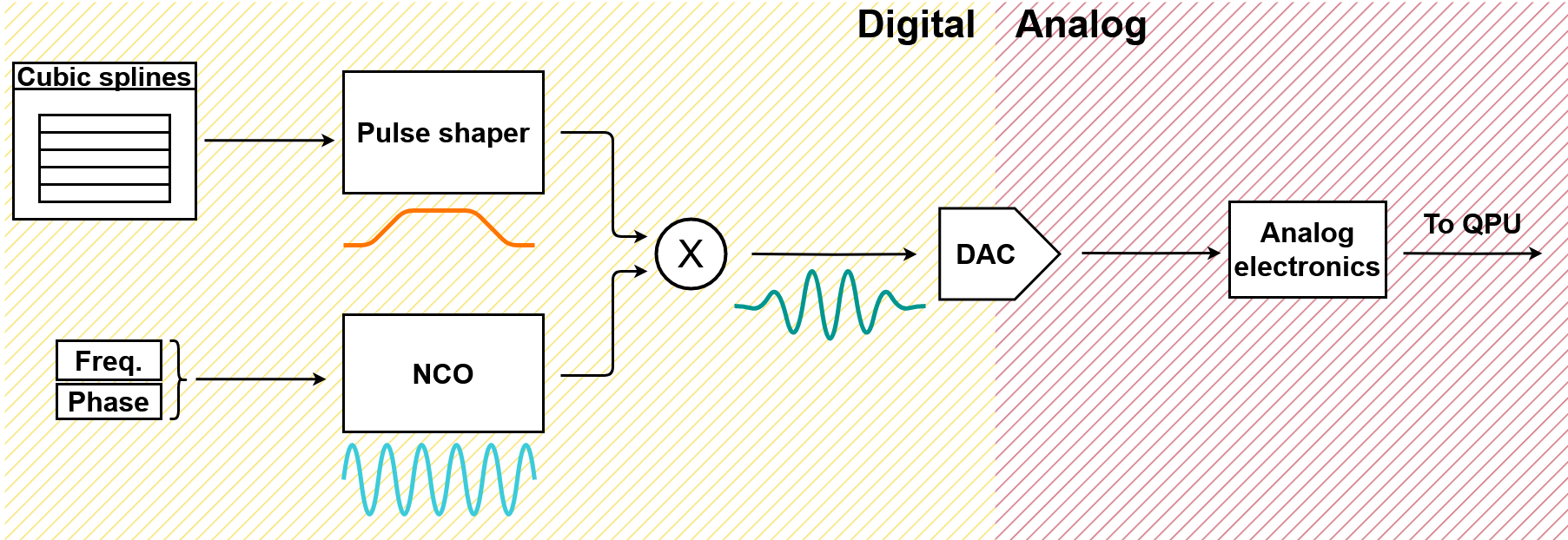}
\caption{Illustrative diagram of Direct Digital Synthesis (DDS) using cubic splines for amplitude modulation.\label{fig:dds}}
\end{figure}

We use Direct Digital Synthesis (DDS) to generate RF pulses from digital circuitry, using FPGA platforms. This is a widely used method in quantum control systems and consists of generating a pure sinusoidal waveform using a Numerically Controlled Oscillator (NCO) modulated in amplitude or phase where the modulation data is pre-stored into memory and multiplied with the NCO output to produce a pulse at base-band frequency \cite{Bowler_paper} \cite{dds_princeton} \cite{DDS_SoC_2020}. This signal is then converted to the analog domain through a Digital-to-Analog Converter (DAC). Additional electronics are used for amplification, filtering and up-conversion (if the desired frequency is outside of the baseband). Figure \ref{fig:dds} illustrates the DDS method specifically applied using compressed envelope data as cubic splines. It should be noted that we perform the up-conversion to microwave frequencies digitally via an NCO tightly integrated with the DAC made possible by the advances in high-bandwidth converter technology over the last decade, largely driven by the telecommunications sector. This removes the need for analog mixers used traditionally, which require calibration and hinder scalability.
In this work, we zoom in on the modulation of the base-band signal, which is the core part of the signal generation chain.

The analytical function for cubic spline polynomials has a non-negligible implementation cost in digital circuitry, mainly due to the multiplication operations involved. In an integer representation, which is more resource-efficient than floating-point, the result requires twice as many bits to represent than the operands. The result is often truncated or rounded to maintain constant multiplier size - particularly in applications requiring cascaded multiplication - which introduces numerical errors when compared to a floating-point representation and needs to be considered.

A coefficient transformation enables the computation of a digitally sampled cubic polynomial recursively as shown in \cite{BowlerSplines2015}. The recursive expression is purely additive, resulting in lower-cost hardware implementation in terms of FPGA resource utilisation and maximum achievable clock frequency.
The output voltage $y$ at sampling time $t$ is defined as:
\begin{equation}
\label{eq:bowler}
    \begin{aligned}
         y(t):=\begin{cases}
            \alpha_0 \hfill  \text{ when $t=0$} \\
            \alpha_{t-1}+\beta_{t-1}+\gamma_{t-1}+\delta_{t-1}  \text{ when $t>0$}
        \end{cases}
    \end{aligned}  
\end{equation}
The coefficients $(\alpha, \beta, \gamma, \delta)$ follow the following recursive definitions:
\begin{equation}
    \begin{aligned}
           \gamma_t= \gamma_{t-1}+\delta_{0} \\
           \beta_t= \beta_{t-1}+\gamma_{t}\\
            \alpha_t= \alpha_{t-1}+\beta_{t} \\
           \delta_t= \delta_0
    \end{aligned}  
\end{equation}
The initial coefficient definitions follow from generalising the recursive formula with respect to cubic polynomial coefficients such that $y(t) =\alpha_t = p^0 + p^1 t + p^2 t^2 + p^3 t^3$ (cf. \eqref{eq:cubic_spline_def}):
\begin{equation}
\label{eq:bowler_coefs}
    \begin{aligned}
            \alpha_0= p^0   \\
           \beta_0= p^1-p^2+p^3\\
           \gamma_0= 2p^2-6p^3\\
           \delta_0= 6p^3
    \end{aligned}  
\end{equation}

The symmetry exploitation is implemented analogously for the coefficients which describe the first $N/2$ segments for a pulse which is segmented into $N$ total segments. Its recursive formula is very similar to that given in Eq. \ref{eq:bowler} and a detailed description is given in App. \ref{sec:app_symmetric}.

\begin{figure}[t!]
\includegraphics[width=0.99\linewidth]{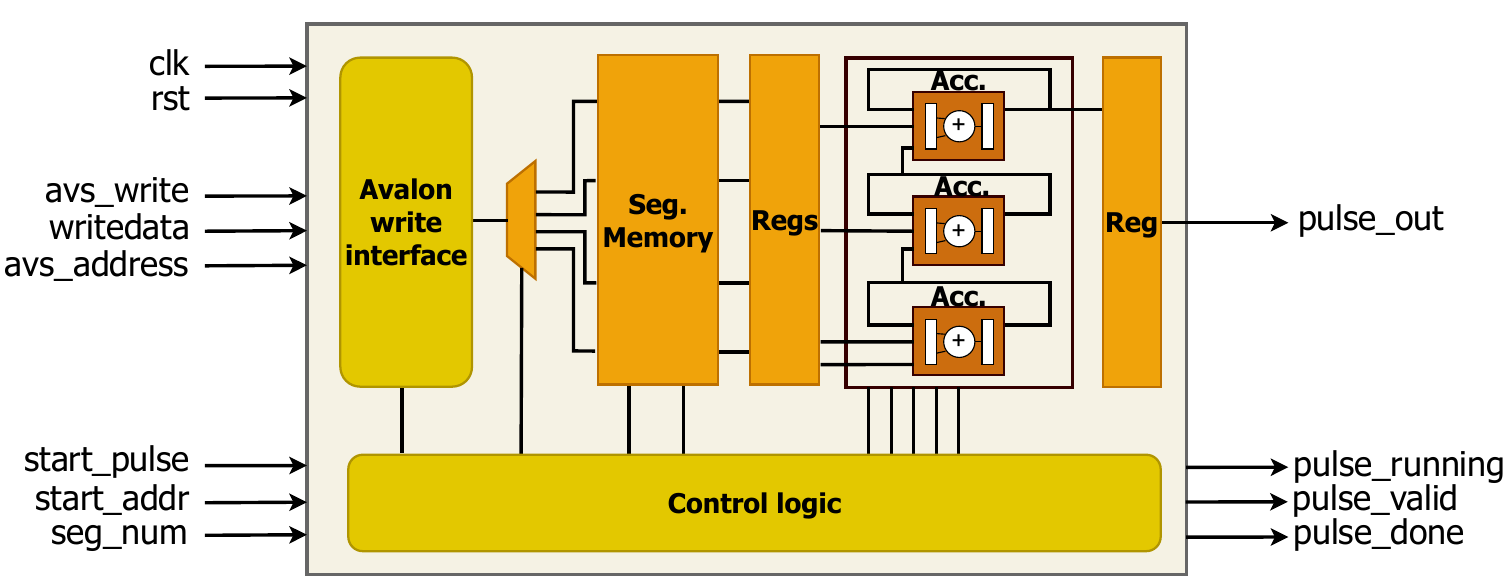}
\caption{Pulse shaper block diagram as implemented on FPGA.\label{fig:pulse_shaper}}
\end{figure}

We implemented the recursive algorithm for cubic splines inside our signal generation IP and deployed it on RFSoC FPGA platforms. Firstly, we performed a quantisation of the modified polynomial coefficients to fixed-point representation for which the bit-width was determined from numerical analysis. This analysis showed that regardless of the chosen bit-width, errors would accumulate inevitably due to the recursive nature of the computation as illustrated in Figure \ref{fig:bowler_quant_errors} of App. \ref{sec:app_quantisation_width}. Increasing the fixed-point precision would increase the memory footprint and in turn, reduce the effective compression factor. Our hardware implementation uses a 36-bit fixed-point representation for the coefficients (i.e. for $\beta$) with 20 bits for the fractional part. This is a compromise between accumulated error and cost for pulses in the range of a few to hundreds of microseconds. It is also the bit precision that was used in the latest implementation of this algorithm in the context of pulse generation for qubit control \cite{BowlerSplines2015}. Some more details are provided in App. \ref{sec:app_quantisation_width}.
The implementation is a pipelined structure where the previous result is added to the new input in each accumulator.
As observed from Eq. \ref{eq:bowler}, $y(t)$ is equal to $\alpha_t$ at every iteration which allows for an implementation with 3 accumulators.
The pipeline produces an output at every clock cycle, maximising throughput. The sampling of input coefficients is handled by a Finite State Machine (FSM) to ensure proper "stitching" between different cubic spline segments in the same pulse, avoiding discontinuities. 
This pipeline can be placed physically on FPGAs as standard Combinational Logic Blocks (CLBs) and registers or on Digital Signal Processing units (DSPs) available on modern FPGA platforms. We have achieved a 575 MHz clock frequency using the latter implementation on an RFSoC Gen-1 chip (XCZU28DR) from AMD/Xilinx. Figure \ref{fig:pulse_shaper} shows the block diagram as implemented on the FPGA. The Avalon interface is a standard protocol for bus data transfers, the Segment Memory is used to store the initial coefficients for each segment and is mapped onto BRAMs, a unit of addressable memory in Xilinx FPGAs, and the DSP48E1 is a dedicated silicon unit supporting up to 48-bit addition. The design uses 84 Look-up Tables (LUTs), 389 registers, 5 DSPs (2 extra used in the control logic), and 4 Block RAMs, all within 1\% of available resources of the target FPGA.

The hardware implementation can be adapted to support symmetry-aware fits, where - in the simplest case of a single exploited symmetry - the latter pulse half is generated from compressed data only available for the earlier pulse half.
Only lightweight modifications are required, compared to the baseline implementation. They mainly consist of re-loading the pipeline with the final coefficients of the first half, introducing a sign inversion, and including control logic to time the switching between computing the recursive function forwards or backwards.

We have also implemented a different version of the hardware which was designed to run at the lower FPGA clock frequency of 250 MHz but with a throughput of 4 samples per clock cycle to increase the sampling rate of pulses to up to 1 GSps allowing to directly generate pulses with a bandwidth of up to 400 MHz.

\subsection{Quantisation-aware fit}
\label{sec:quantised_fit}

\begin{figure*}[t!]\includegraphics[width=\textwidth]{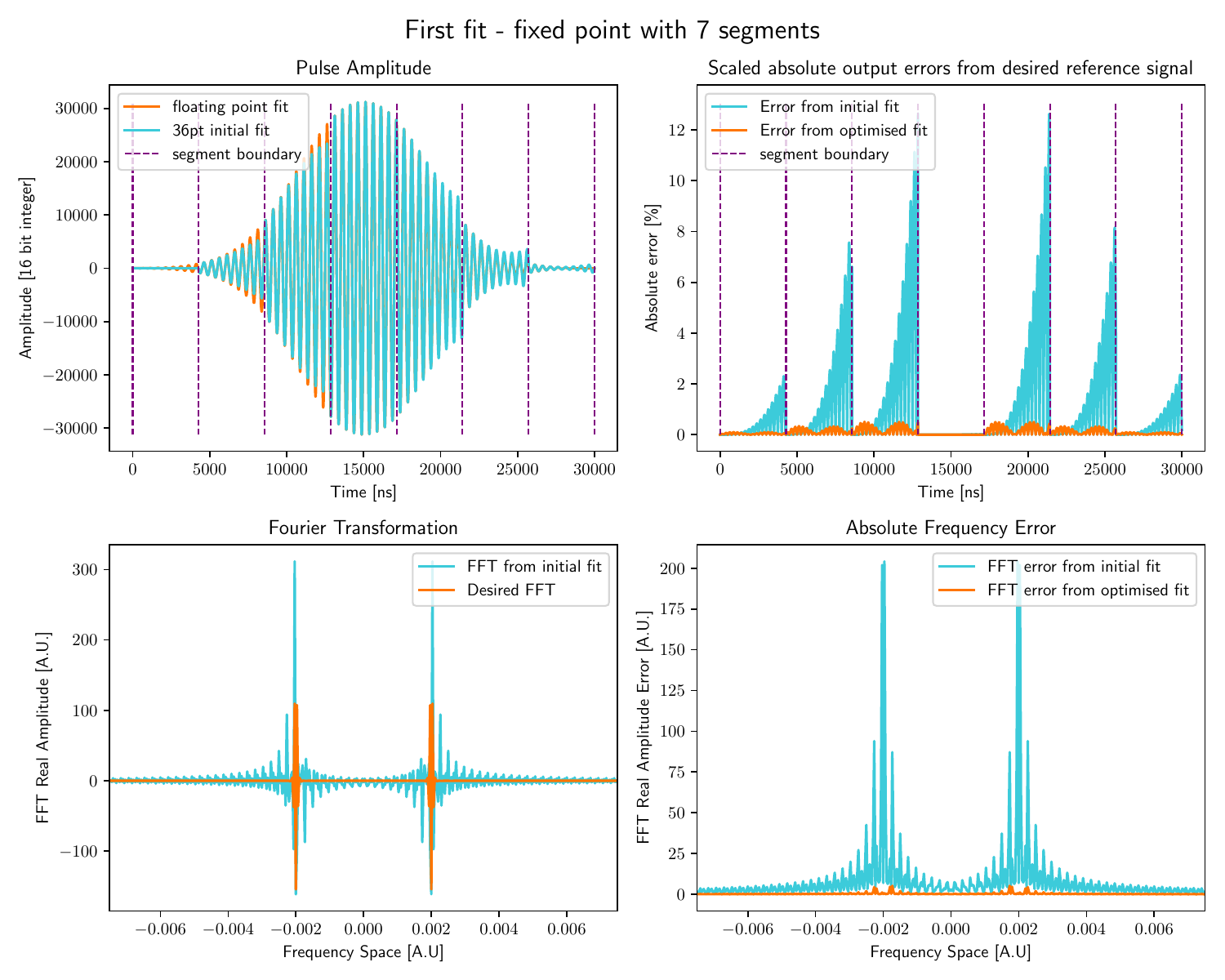}
\caption{We show the effect of the quantisation-aware fitting method for a Gaussian pulse of length $30000$ partitioned into $7$ segments and multiplied with an NCO frequency of $f=2$MHz for illustration purposes. Top-left: Comparison of the ideal floating point fit of the desired Gaussian amplitude envelope ($y_{ref}(t)=\sin(2 \pi ft)e^{-(t-15000)^2/(1.6e^7)}$) to the quantised 36pt initial fit. Top-right: Absolute percentage error of the quantised fits with respect to the desired output. The optimised fitting method reduces the error by an order of magnitude. Bottom-left: Fourier Transformation of the desired signal and the initial fit (where x=0 corresponds to the centre frequency of $2$ MHz). Bottom-right: Absolute error in Fourier Transform between quantised fits and the desired signal. The optimised fit again improves the error by an order of magnitude. The cubic polynomial coefficients for two example segments are shown for reference in App. \ref{sec:examples} \label{fig:optimised_fit}}
\end{figure*}

Typically, gradient-based minimisation algorithms such as the Python Scipy CubicSpline \cite{2020SciPy-NMeth} module is used to fit functions to a representation with cubic spline segments in floating point arithmetic, i.e. operations and coefficients can be carried out and stored at a high level of accuracy (typically 64 bit). Such algorithms are very mature, they are fast, well-understood, and well-documented. The hardware implementation of the recursive algorithm described in Sec. \ref{sec:fpga} used to represent cubic splines is shown in Fig. \ref{fig:pulse_shaper}. Note that the representation of the linear combinations of cubic polynomial coefficients $ \alpha, \beta, \gamma \text{ and } \delta$ is fixed point, not floating point. This implies that there are rounding errors arising from the quantisation, depending on the number of fractional bits used in the fixed point representation (Figure \ref{fig:optimised_fit} shows 36-bit total word length with 20-bit fractional accuracy). This leads to the following error accumulation when decompressing the pulse envelope:
\begin{equation}
\begin{aligned}
\label{eq:error}
    \epsilon_{seg}[n]&=\epsilon_{\alpha_0}+n\epsilon_{\beta_0}+\frac{n(n+1)}{2}\epsilon_{\gamma_0}+\frac{n(n+1)(n+2)}{6}\epsilon_{\delta_0}
\end{aligned}
\end{equation}
where the $\epsilon_{i}$ denote the quantisation errors associated with the $i$-th coefficient described in Eq. \eqref{eq:bowler_coefs}, i.e. when truncating any bits that lie outside the realm of the fixed point quantisation and $n$ denotes the number of clocks (i.e. addition) cycles per segment. The accumulated errors depend strongly on the number of sum operations $n$ performed per segment. A larger number of additions per segment leads to better compression factors but also to an increase in the accumulated errors which are $O(n^3)$. For sufficiently large $n$, $\epsilon_{\delta}$ starts dominating, and it typically becomes impossible to output non-trivial smooth pulse envelopes. We find that for a typical example of a Gaussian pulse,  with a 36-bit representation, for segment lengths exceeding $1000$ rounds of addition, the accumulated errors become severe as shown in Fig. \ref{fig:optimised_fit} and exceed the tolerances required by most quantum computers (cf. Sec. \ref{sec:benchmarking}). Note that 2000 rounds of addition corresponds to a segment length of 2000 ns at 1 GSps, or 4000 ns at 500 MSps, typical sampling rates used for pulse generation in quantum computing, which are pulse lengths realistically required for atomic qubits.
These errors are particularly problematic as the kinks create undesired frequency components which adversely affect the fidelity of quantum operations. One could reduce the individual segment length and use more segments to represent the pulse, but this reduces the compression factor and increases the memory load, which is undesirable in memory-limited environments. Increasing the length of the bit representation of the coefficients also decreases the compression factor. Moreover, if the number of samples per segment is large enough, any fixed point quantisation will introduce significant errors, particularly towards the end of an individual segment, due to the cubic dependence of the segment error function (see Eq. \ref{eq:error} and cf. App. \ref{sec:app_quantisation_width} for more details).

We have developed a strategy to mitigate this, which relies on a two-stage fitting process: 
\begin{enumerate}
    \item Fit the desired function with a standard cubic spline fitting algorithm as explained in Sec. \ref{sec:floating_point_fit} which relies on floating point arithmetic. 
    \item Use the coefficients obtained in the first fitting round as the starting point for a second fitting algorithm which incorporates the fixed-point accuracy of the recursive algorithm described in Eq. \ref{eq:bowler}.
\end{enumerate}
The second step requires an evaluation of the error between a floating point implementation of equation \ref {eq:bowler} and a quantised implementation to construct an adequate cost function for the optimisation algorithm. A brute force approach would be to directly implement the recursive algorithm described in Eq. \ref{eq:bowler}, but we were unable to find a computationally efficient way of doing so. Instead, we used the error function described in Eq.\ref{eq:error} which resulted in drastic speed improvements.


The constant coefficient $\alpha_0=p^0$ is kept fixed for all segments, but the coefficients $\beta_0, \gamma_0, \delta_0$ for each segment are treated as variables which are to be optimised as ($\beta^{*}_0, \gamma^{*}_0, \delta^{*}_0$). The cost function per segment $\mathcal{E}$ for the quantisation-aware fitting method we used is given by:
\begin{equation}
\begin{aligned}
    \mathcal{E}_{i}=\sum_n^{N}(y_n-\left\lfloor y_n + \epsilon^*_{seg}[n]\right\rfloor)^2,
\end{aligned}
\end{equation}
where $n$ describes the number of additions performed per segment, $y_n$ describes the desired pulse amplitude and $\epsilon^*_{seg}$ the quantisation error associated with the optimised coefficients $(\beta^{*}_0, \gamma^{*}_0, \delta^{*}_0)$ with respect to the floating point coefficients determined in the first fit $(\beta_0, \gamma_0, \delta_0)$. \footnote{We explored alterations to the cost function above, in particular adding the maximum error to the average error over a segment with equal weighting, which marginally improved the accuracy of the optimised fit for some pulses but not universally.} The least squares error can be minimised with a variety of non-gradient descent-based optimisation methods. A non-exhaustive list of optimisation approaches to this problem is: simulated annealing, pattern search, grid search, particle swarm optimisation, Bayesian optimisation, or particle evolution. In our implementation, the latter was found to have a particularly good trade-off between speed and accuracy. For the examples shown here the runtime of the algorithm was on the order of minutes on a standard laptop (Intel Core i7-1185G7 @ 3 GHz with 8 logical cores).

\section{Benchmarking}
\label{sec:benchmarking}
Neutral atoms have emerged as promising platforms for quantum computing, with thousands of qubits \cite{manetsch2024tweezerarray6100highly} with high fidelity parallel entangling gates \cite{Evered2023, Bluvstein2023} and long coherence times. In what follows, we will apply the quantisation-aware cubic spline fitting routine to simulations of a neutral atom single-qubit gate as well as atom transport. An explicit comparison with the floating point fit as well as the AWG implementation is provided and the optimised fit is shown to exhibit a particularly favourable trade-off between the fidelity of the operation and the memory required to store the waveform.
\subsection{Coherent Quantum Population Transfer in $^{87}\mathrm{Rb}$ Atom}


To benchmark the fitting methods against the state of the art, a simple model of full coherent quantum population transfer (a particular variant of a single-qubit gate) is considered, and the effect of the pulse envelope shaping method on the fidelity is calculated. Single-qubit gates in atomic quantum computing architectures can be implemented in multiple ways. STImulated Raman Adiabatic Passage (STIRAP) \cite{STIRAP_review} is a technique commonly used for adiabatically transferring population between two ground states, which encode either of the qubit basis states by using a Hamiltonian eigenstate which has no contribution from the excited state. This typically requires two laser pulses that have equal detunings for a common excited state to establish a coupling between the two ground states. In the simplest case, invoking the rotating wave approximation (cf. \cite{vasilev2009} and \cite{shore_atomic}), the Hamiltonian describing this interaction in the $(\ket{0}, \ket{1}, \ket{e})$ basis reads:
    \begin{equation}
        \mathbf{H}=\frac{\hbar}{2}\begin{bmatrix}
                                0 & 0 & -\Omega_1(t)\\
                               0 & 0 & -\Omega_2(t) \\
                               -\Omega_1(t) & -\Omega_2(t) & 2\delta
                                \end{bmatrix},
    \end{equation}
where we describe the time-dependent Rabi frequencies $\Omega_1(t)$ and $\Omega_2(t)$, as well as the detuning from the excited state $\delta$. The Hamiltonian contains a dark eigenstate which has no contribution from the excited state:
\begin{equation}
    \ket{\psi_d}=\cos(\phi(t))\ket{0}-\sin(\phi(t))\ket{1},
\end{equation}
where the mixing angle reads $\phi(t)=\arctan(\frac{\Omega_1(t)}{\Omega_2(t)})$. In order to perform a single-qubit rotation with the dark state $\ket{\psi_d}$ which contains no contribution from $\ket{e}$, amplitude \footnote{In some cases phase modulation can be used to implement amplitude modulation \cite{scaleable_driving}.} modulation of the pulses  $\Omega_{1/2}(t) $ is required to realise high-fidelity quantum population transfer and single-qubit rotations. We benchmark our optimised curve fitting algorithm on a typical pulse sequence used to realise single-qubit gates in $^{87}\mathrm{Rb}$ where the two qubit states are given by the hyperfine clock states $F=2, m_F=0$ and $F=1, m_F=0$  which, at a particular external magnetic field, are to first order insensitive to fluctuations in that external magnetic field. $\Omega_1$ and $\Omega_2$ are both $\sigma^{-}$ polarised. This single-qubit gate implementation with two beams was experimentally demonstrated with Rb atoms in \cite{Grangier_2007, saffman2005}, as well as with microwaves \cite{microwave_magicwavelength, Beterov_2021} and with a single laser beam which is far detuned from the transitions \cite{scaleable_driving}.

\begin{figure}[ht!]\includegraphics[width=0.99\linewidth]{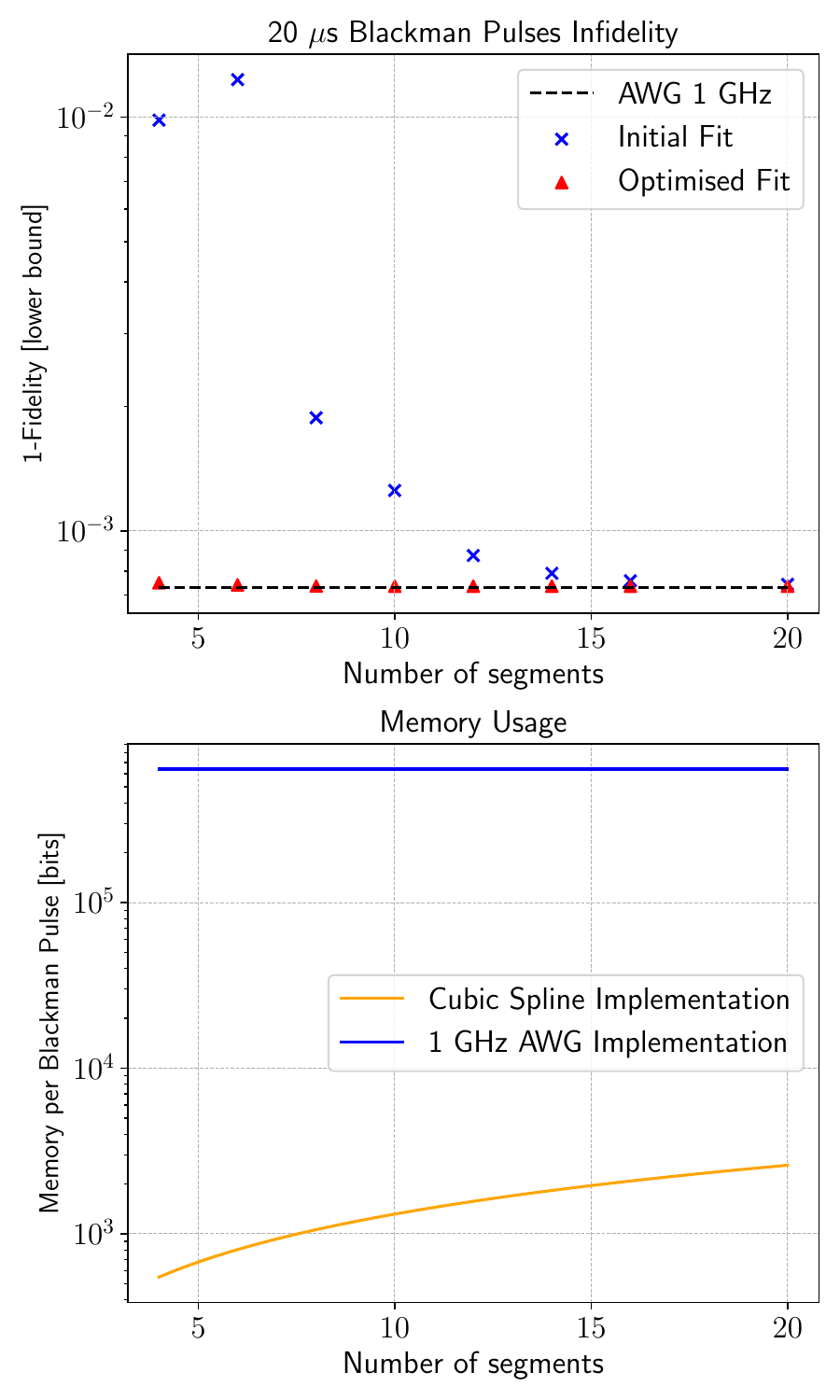} \caption{We consider two fixed length 20 $\mu$s Blackman pulses used to implement an $X(\pi)$ gate and compare the fidelity (top) as well as the memory usage to store the pulse (bottom) between the optimised fit, the floating point fit and the AWG implementation. The top plot shows that as the number of segments increases, the optimised fitting fidelity no longer differs from the fidelity obtained with the initial fit, however the point where the optimised fit reaches a comparable fidelity to the AWG is already at $\approx6$ segments (compared to $\approx20$) which implies a significantly lower memory overhead. This is exhibited in the bottom plot where one can see that we obtain almost three orders of magnitude of compression for $6$ segments when comparing the cubic spline and AWG implementations. \label{fig:2mu_pulse_segs}}
\end{figure}
We consider a simple model of the atom taken from reference \cite{Ernst_2023} and isolate the error induced by the imperfect amplitude modulation of the envelope. Other error sources such as phase/frequency fluctuations, laser power fluctuations, magnetic field fluctuations, other imperfections in the electronic signal chain, or laser frequency fluctuations, optical aberrations, and atomic motion are completely ignored. It shall be stressed that in the small-error regime where quantum computers typically operate, the errors are to first order linearly additive, which justifies this approximation to benchmark different fitting methods. The only errors arise from scattering, and we neglect other error sources, as we are interested solely in the error arising from imperfections in the amplitude modulation. These are significant enough that they can be clearly shown in Fig.
\ref{fig:2mu_pulse_segs}.  Given the highly idealised nature of the model, all fidelities in the results which follow are strictly upper bounds in achievable experimental fidelity, given the onset of a variety of other possible imperfections or errors. In what follows, we initialise the atom in $\ket{1}$ and implement an $X(\pi)$ rotation \footnote{One could extend this to arbitrary $X(\theta)$ rotations but we focus on one particular $\theta=\pi$ rotation for simplicity.}. The time-dependent amplitude of $\Omega_{1 / 2}$ are both Blackman envelopes \footnote{There are asymmetric time-dependent amplitudes \cite{optimal_stirap} which lead to higher population transfer fidelities but a Blackman ramp is considered here for simplicity.} with a fixed delay factor of $0.3L$ where $L$ are the lengths of both $\Omega_1$ and $\Omega_2$, such that the root-mean-square Rabi frequency $\sqrt{\Omega_1^2(t) + \Omega_2^2(t)}$ is approximately constant. The total gate time is thus $1.3 L$, where $L=20$ $\mu$s in Fig. \ref{fig:2mu_pulse_segs}. The peak Rabi frequency was fixed at $5$ MHz with a fixed detuning of $-100$ MHz from the $F'=1$ level, this implies that longer pulses have a larger pulse area and thereafter a higher fidelity for this particular X rotation. The external magnetic field is fixed at $\approx 0.7$ mT to induce a fixed ground state Zeeman splitting of $5$ MHz. These are all feasible experimental parameters.

The results show that in this highly idealised simulation of a single $^{87}\mathrm{Rb}$ atom the quantisation-aware fitting routine has a noticeable effect on the fidelity and leads to a negligible reduction in fidelity comparison to the brute-force AWG implementation, particularly for longer pulses with fewer numbers of segments. In Fig. \ref{fig:2mu_pulse_segs} one can observe that for fixed-length pulses the quantisation-aware fitting routine is particularly significant up to around $14$ segments, but its effects trail off when individual pulse segments contain fewer than $1000$ samples (when the number of segments exceeds $20$) \footnote{The cubic spline block is assumed to be implemented at an output sampling rate of $1$ GHz.}. This figure shows that to achieve a maximal fidelity upper bound (given by the AWG pulse representation) one only requires $6$ (with optimised fitting) compared to $20$ (without optimised fitting) segments, which more than halves the required memory to store the pulse information compared to the unoptimised fitting and is a compression ratio of $\approx 796$ when comparing the brute force AWG approach and the cubic spline with an optimised fit with $6$ segments. 


\subsection{Atom Transport}
\begin{figure}[ht!]\includegraphics[width=0.99\linewidth]{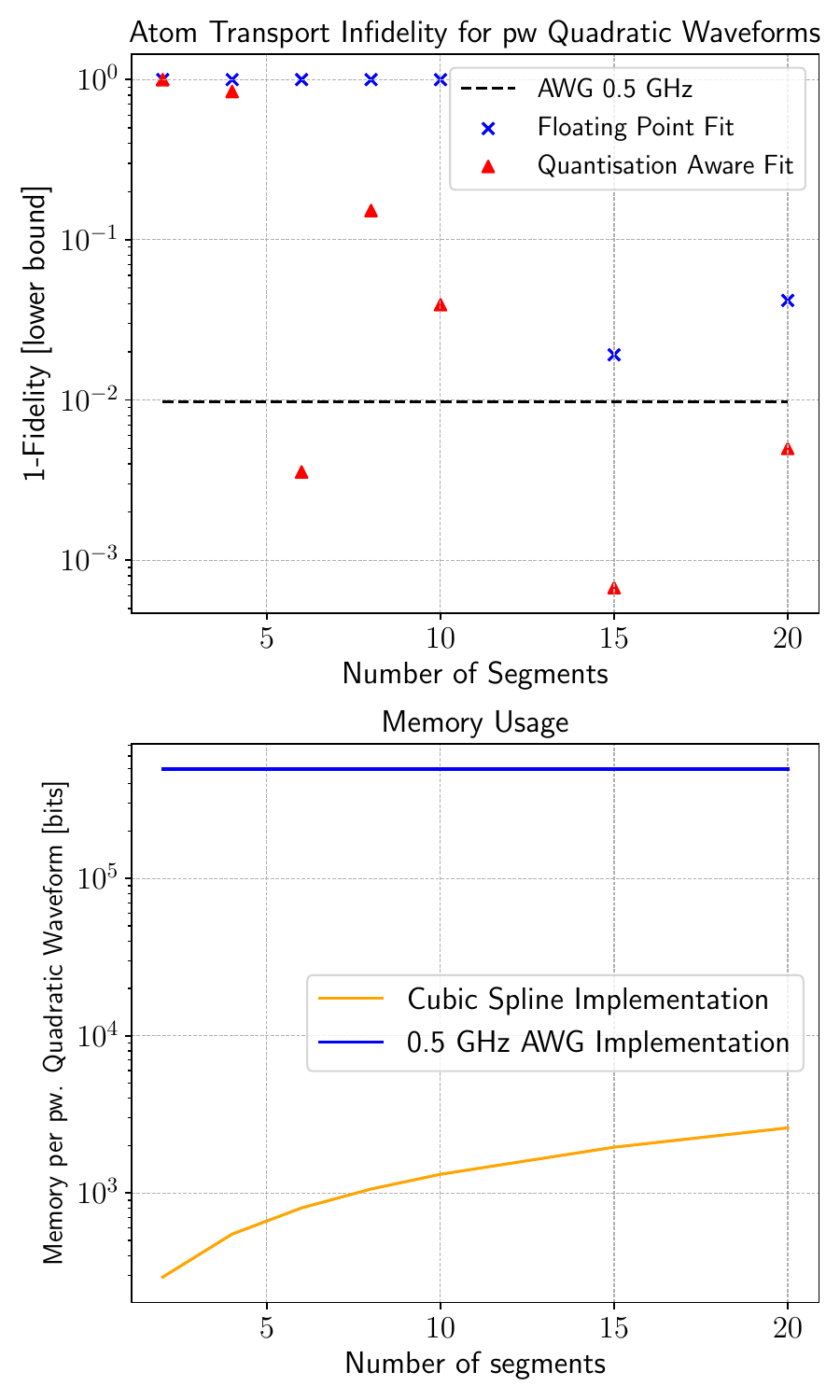}
\caption{
We compare fidelities (top) across the floating point fit, the optimised fit, and the AWG implementation for an atom transport with a piecewise quadratic pulse (cf. Eq. \ref{eq:pw_qu} with a length of $31 \mu$s and a trap depth of $20 \mu$K) and also compare their respective memory usage (bottom). The top plot shows that for a small number of segments the cubic polynomial representations are insufficient for high-fidelity transport even with the optimised fitting because small pulse kinks lead to immediate loss of the atom from the trap. At larger segment numbers we can simulate atom transport with extremely high fidelities and much more modest memory costs. The improvement in fidelity from floating point to quantisation-aware fit can be as large as two orders of magnitude. The fidelity in fact can even exceed that attained with the AWG implementation, as the piecewise quadratic pulse used here is not the optimal waveform for atom transport \cite{pagano2024optimalcontroltransportneutral}. The bottom plot shows similar results to those obtained in Fig. \ref{fig:2mu_pulse_segs} and we obtain almost three orders of magnitude of compression at 6 segments when comparing the optimised fit to the AWG implementation. \label{fig:shuttling_waveforms}}
\end{figure}

 High-fidelity transport of neutral atoms is vital for the initialisation of stochastically loaded atomic arrays for analog simulation and computing, as well as for the dynamic rewiring of qubit connectivity. Atoms are typically loaded into an array of static tweezers created by spatial light modulators, however, the loading probability of getting exactly one atom in an individual trap is $<1$ \cite{loading_eff}, and often closer to 50\%, and to form a desired topology atoms have to be moved to the desired trapping sites. This is achieved by overlaying static traps with another set of dynamic traps controlled by acousto-optic deflectors (AOD). Precise control of the frequency of the drives induces spatial movement of trapped atoms\cite{Schymik2020, Endres2016, barredo2016}.  Analogously, the qubit connectivity can be modified mid-algorithm \cite{Bluvstein2022, Bluvstein2023}.
The movement of the atoms must be carried out at high fidelities and speeds to prevent loss and to avoid introducing an unnecessary time overhead. The speed at which atoms can be moved is bounded by the quantum speed limit, which for a distance of $3$ $\mu m$ is found to be around $10$ $\mu s$ \cite{pagano2024optimalcontroltransportneutral}.
Given the desire to move thousands of atoms with multiple tones over many trapping sites, which requires precise control over the frequency and amplitude of the transport waveforms, we investigated pulse compression to reduce the hardware overhead required for the memory-efficient representation of such signals.

The description of the transport simulation follows closely from the description in \cite{pagano2024optimalcontroltransportneutral} and the same code is used for a numerically realistic simulation of atom transport, where we consider the motion of a single $^{88}\mathrm{Sr}$ atom over a distance of $3$ $\mu$m which separates adjacent trapping sites (this is equivalent to a $1$ MHz frequency modulation on the trapping waveform). Every atom is considered to be in a thermal state \cite{pagano2024optimalcontroltransportneutral}:
\begin{equation}
\begin{aligned}
   \rho(T) &= \sum_{i=1} p_{i} \ket{\psi_{i}}\bra{\psi_{i}}
   \label{eq:thermal-state}
\end{aligned}
\end{equation}
where $p_{i} (T) = \mathrm{e}^{-E_i/k_b T}$ are the Boltzmann weights at temperature $T$ (for more details cf. \cite{pagano2024optimalcontroltransportneutral} where the truncation of this sum is discussed). The optical tweezer is modelled as a Gaussian potential \cite{pagano2024optimalcontroltransportneutral}:
\begin{align}
    V(t) = U_d \mathrm{e}^{ \frac{- 2 r(t)^2}{w_0^2}  } \, ,
    \label{eq:gaussian-trap}
\end{align}
where $U_d$ is the trap depth, $w_0$ the beam waist and $r$ the time-dependent centre of the trap \cite{Endres2016}. In the optical setup there is correspondence between the frequency of the driving tone and the position of the tweezer $r(t)$ and the tuning of $f(t)$ creates a time-dependent motion of the trapping centre $r(t)$. Without further optimising the shape of the frequency chirp \footnote{For simplicity we picked a simple pulse representation, but there are higher-order polynomials or numerical solutions determined by optimal control methods (cf. \cite{pagano2024optimalcontroltransportneutral} for more details) which lead to higher transport fidelities.} We consider an extremely simple piecewise quadratic pulse with time-dependent frequency:
\begin{equation}
f_{\mathrm{pq}}(t) =
\left\{
\begin{aligned}
   & f_0 + 2 \frac{f_\mathrm{f} - f_0}{t_\mathrm{p}^2} t^2 & \quad & t \leq t_\mathrm{p}/2 \\
   & f_\mathrm{f} - 2 \frac{f_\mathrm{p} - f_0}{t_\mathrm{p}^2} (t-t_\mathrm{p})^2 & \quad & t > t_\mathrm{p}/2
\end{aligned} \right.
\label{eq:pw_qu}
\end{equation}
 where $t_\mathrm{p}=31$ $\mu$s is the duration of the pulse and $f_f, f_0$ are the final and initial frequencies respectively, such that $f_f-f_0=1$ MHz which corresponds to an adjacent trapping site. We benchmark the fidelity of the atom transport using different signal representations with an AWG as well as the aforementioned floating and quantisation-aware cubic spline fits. For this purpose, we follow \cite{pagano2024optimalcontroltransportneutral} and consider the infidelity between the final density after applying the shuttling waveform $\rho_f(r)$ and the initial density matrix shifted to the final spatial position $\rho_0(r_f)$:
 \begin{align}
    J (r) = 1 - {\mathrm{Tr} \left(\sqrt{ \sqrt{\rho_\mathrm{f}(r)} \rho_0 (r_\mathrm{f}) \sqrt{\rho_\mathrm{f}(r)}}\right)}^2
    \label{eq:figure-of-merit}
\end{align}
This is averaged over $10 \mu$s due to oscillatory behaviour after the atom transport \cite{pagano2024optimalcontroltransportneutral} which confounds the investigation of the effect of the waveform on the post-transport fidelity. It shall be stressed that we isolate the effect of a memory-efficient representation of the transport waveform which has a time-dependent frequency and does not consider any other error sources which induce noise on the trap depth, position, and waist arising from laser noise, fibre coupling, finite AOD rise time and other imperfections which are prominent in physical experiments. As in the previous section, we highlight that in the small-error regime where quantum computers typically operate, the errors are to first order linearly additive, which justifies this approximation. The effects of the imperfect frequency modulation are also significant and visibly impact transport infidelities by orders of magnitude, as shown in Fig. \ref{fig:shuttling_waveforms}. We consider a representation of the piecewise quadratic pulse with cubic splines using the floating point fit and quantisation-aware point fit described in Sec. \ref{sec:quantised_fit} and compare this to a brute force AWG output at a sampling rate of $500$ MHz (this is more realistic for longer shuttling waveforms of up to several hundreds of microseconds where higher sampling rates are not necessary) for a $31$ $\mu$s piecewise quadratic pulse described in Eq. \ref{eq:pw_qu}.
The highly idealised numerical simulation results for atom transport in Fig. \ref{fig:shuttling_waveforms} show that in the low segment limit, a cubic spline fit cannot achieve high-fidelity atom transport even with the optimised fit. However, for the case of $6$ segments, the optimised fit shows over two orders of magnitude reduction in infidelity compared to the floating point fit and attains an even lower fidelity than the AWG representation at a memory compression of $\approx 620$. For increasing segment numbers the results are more erratic due to the small variations in the representation of the piecewise quadratic function, but consistently the results show that the optimised fit significantly reduces atom transport infidelity, often by more than one order of magnitude. Similar results were attained with different trap depths. In addition, although we do not show this, moving tweezers over larger distances in large atom arrays requires longer waveforms of up to hundreds of microseconds with higher memory overheads of up to $\approx O(10^6)$ bits when storing the frequency envelopes directly in the memory of an AWG. This further motivates the use of the optimised compression techniques outlined in this paper.

It should be noted that here the piecewise quadratic pulse represented by an AWG does not constitute an upper bound on transport fidelity, as it was shown in Ref. \cite{pagano2024optimalcontroltransportneutral} that more optimal waveforms can be determined with quantum control methods. Indeed, for a large enough number of segments we found that waveforms represented with cubic splines in some cases achieve higher transport fidelities as shown in Fig. \ref{fig:shuttling_waveforms}, however, this is purely by chance and not by design. We leave the determination of an optimal transport waveform (i.e. not piecewise quadratic) which can be efficiently represented with cubic splines as a subject for future work.

\vfill

\section{Outlook}
We have shown how radio-frequency signals, particularly relevant for atomic quantum computing systems, can be compressed with cubic splines and demonstrated an efficient implementation on an FPGA with high output sampling rates of up to $1$ GHz by using a series of recursive sums to generate output samples. In particular, two memory compression methods, one that exploits pulse symmetry and one that relies on a hardware-aware, fixed-point fit were described to improve pulse representation without compromising on the achievable compression factor. The merits of the optimised fitting routine were demonstrated in the context of a single-qubit gate simulation on a neutral atom quantum device as well as for atom transport with optical tweezers, and we showed significant improvements in fidelity over the floating point fit. In particular, for longer pulses or fewer segments, we could show close to no reduction in fidelity to the state-of-the-art AWG method and compression factors exceeding two orders of magnitude.

Although in this paper we focused on applications in neutral atom quantum devices, the pulse engineering techniques are platform-agnostic. Analogously to applying the technique to neutral atom transport, it is likely to find application to trapped-ion transport waveforms, where time-varying electric potentials are applied to multiple ion-trap electrodes to move ions between zones. This may be especially relevant for fast, high-fidelity transport schemes making use of complex signals, such as invariant-based engineering or shortcuts to adiabaticity \cite{fast_shuttling}. 

In contrast to atomic qubits, the pulse duration for single- and two-qubit gates in superconducting quantum computing architectures is typically on the order of tens to hundreds of nanoseconds, well within the coherence times of these qubits\cite{RevModPhys.93.025005}. The bandwidth of electronic devices is typically limited to a few GHz, so most signals only require on the order of hundreds of samples to represent pulses at this bandwidth. Pulse shapes can vary dramatically: commonly used DRAG gates require smooth pulses \cite{drag}, whereas more recent solutions leading to higher fidelity use non-smooth pulses obtained by reinforcement learning methods \cite{rl_superconducting}. The time scales and the non-standard envelope shapes limit the potential of pulse compression techniques, and therefore we focus on applications in atomic quantum systems. However, there might be applications for higher-bandwidth electronic signal generation. Furthermore, there might also be applications in other material platforms such as, but not limited to, quantum dots or colour centres, but we have not performed a comprehensive study of the control requirements.

In general, we hope this paper motivates more work into resource analysis and optimisation for future-proof quantum control systems. Memory is likely a limited resource for compact, integrated control systems, and this work provides some techniques which can be used to minimise memory usage without compromising pulse representation and fidelity of quantum operations.
\section{Acknowledgements}
J.O.E. gratefully acknowledges Alice Pagano for useful discussions about the atom transport simulation.
\bibliography{bibliography}
\section{Appendix}
\subsection{Knot Positioning}
\label{sec:app_knots}
The increase in memory load with the number of segments motivates an optimisation of the position of the segment boundaries, as it can noticeably improve the fitting accuracy without increasing the memory load. As it turns out, the global optimisation of the knot positions is a computationally complex non-linear optimisation problem. Recently, some progress has been made in formulating it as a mixed-integer quadratically constrained problem, as well as a branch and bound algorithm \cite{global_splines}. Previously, only locally optimal optimisation methods had been demonstrated \cite{splines_better, deBoor1973_splines} for the positioning of the segment boundaries. Even though recent results show promise, the computation times for longer pulses with lots of points to fit to are fairly long (order of hours for 6 knots with a few hundred data points on Intel Core i7-1185G7 @ 3 GHz with 8 logical cores) to render this solution practical on our standard laptops. Additionally, the improvements in fitting accuracy for typical pulse shapes used in quantum computing are lower than that gained by exploiting pulse symmetry or fitting per the FPGA level accuracy described in Sec. \ref{sec:quantised_fit}. However, for some particularly critical applications where this optimisation only has to be done once, it could be worth it. 
\subsection{Choice of quantisation width}
\label{sec:app_quantisation_width}

\begin{figure}[ht!]\includegraphics[width=0.99\linewidth]{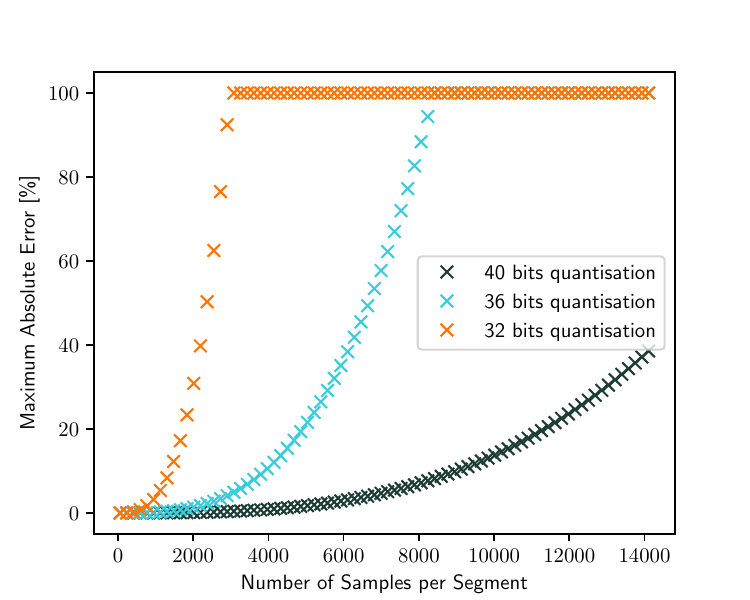}
\caption{We plot the pulse representation error (cf. Eq. \ref{eq:error}) associated with the truncation of $\beta, \gamma$ and $\delta$ at various quantisation widths for an example segment which is also plotted in Fig. \ref{fig:optimised_fit}. It is clear that the errors are reduced by increasing the quantisation width, but due to the cubic nature of the error function, the errors will always persist and blow up for larger sample numbers. $\epsilon_0=\epsilon_{\alpha_0}$ which is the error associated with the truncation of $\alpha$ which does not contribute to the cumulative errors. It is evident that for enough samples, the error becomes extremely noticeable and exceeds up to $0.3$ with a 32-bit width on a $[0,1]$ output scale. \label{fig:bowler_quant_errors}}
\end{figure}

Our analysis is mostly based on 36-bit precision to reflect initial hardware implementations in previous work and at Riverlane, where 32-bit representations would produce large errors. In this particular case, $20$ bits are reserved for the fractional part, $15$ for the integer part, and $1$ for the sign. All computation is done using $36$ bits, but the final result is truncated to $16$ bits. This is because high-speed RF DACs typically have a resolution between $12$ or $16$ bits, hence the least significant bits are ignored. Since $\alpha_0$ corresponds to the first output and is only used once in the recursive algorithm, it only requires $16$ bits of precision saving $20$ bits of memory.

In general, a larger number of samples per segment requires longer bit widths of the coefficients, but any fixed point representation will lead to cumulative errors if the number of samples per segment is large enough. This is exemplified in Fig. \ref{fig:bowler_quant_errors}. Thereafter, it doesn't matter how accurate the fixed point representation is: if the number of samples per segment is large enough, a quantisation-aware fit will substantially improve the pulse representation. Alternatively, one can consider the benefit of the quantised-aware fit on the compression factor, in that the pulse-aware fit allows to achieve the same accuracy as the standard fit for any pulse, but with lower bit-precision.

With the improvements proposed in this work, future hardware implementations can use reduced bit precision and further improve compression factors, with 32 being an optimal width for memory accesses.

\subsection{Alterations to FPGA design to support symmetry}
\label{sec:app_symmetric}
\begin{figure*}[t!]\includegraphics[width=1\linewidth]{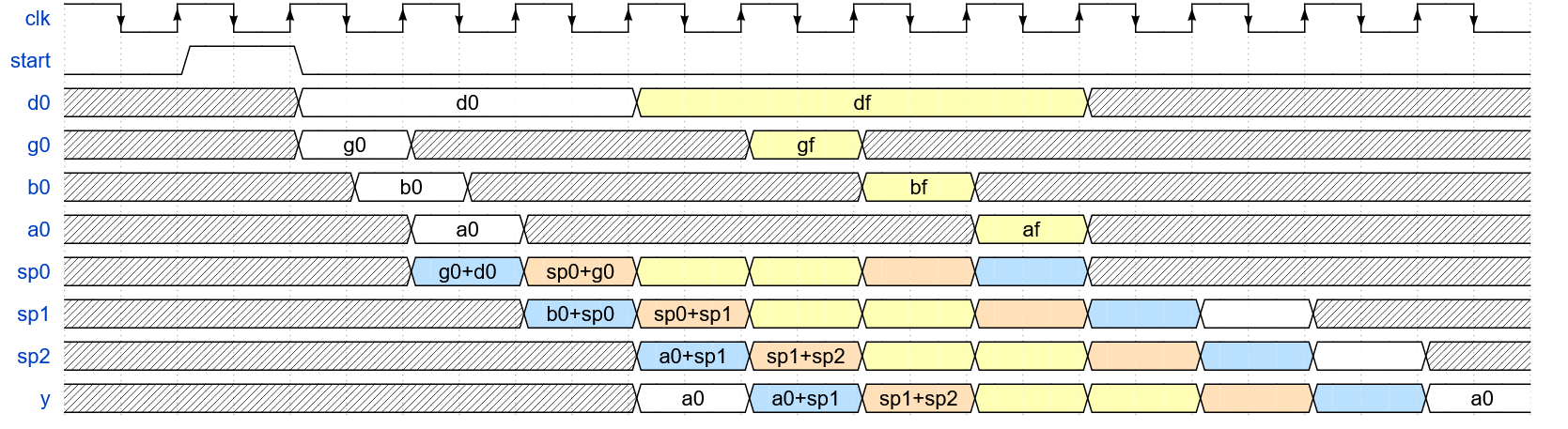}
\caption{Timing diagram for 8 sample symmetric segments. It represents the value of registers at every unit of the block's synchronous clock signal. The data is sampled at every positive edge of $clk$. The $start$ signal is the trigger to start the envelope generation. Labels $a0$, $b0$, $g0$, $d0$ are for the registers storing the coefficients for the first iteration of the recursive algorithm. Similarly, $af$, $bf$, $gf$, $df$ are the initial coefficient values for the symmetric part of the pulse as in Equation \ref{eq:bowler_sym}. $sp0$ to $sp2$ are the pipeline registers storing intermediate accumulated results and $y$ is the output.\label{fig:timing_symmetric_pulses}}
\end{figure*}

The symmetry exploitative recursive formula reads:
\begin{equation}
\label{eq:bowler_sym}
    \begin{aligned}
         y(t):=\begin{cases}
             \alpha'_0 \hfill  \text{ when $t=0$} \\
            \alpha'_{t}+\beta'_{t}+\gamma'_{t}+\delta'_{t}  \text{ when $t>0$}
        \end{cases}
    \end{aligned}  
\end{equation}
The primed coefficients $(\alpha', \beta', \gamma', \delta')$ follow the following recursive definitions:
\begin{equation}
    \begin{aligned}
            \alpha'_t= \alpha'_{t-1}-\beta'_{t-1}     \\
           \beta'_t= \beta'_{t-1}-\gamma'_{t-1}\\
           \gamma'_t= \gamma'_{t-1}-\delta'_{t-1} \\
           \delta'_t= \delta'_0
    \end{aligned}  
\end{equation}
The initial coefficients are the final coefficients of the symmetry-related segment (unprimed coefficients) and read:
\begin{equation}
    \begin{aligned}
            \alpha'_0= \alpha_{t_f}   \\
           \beta'_0= \beta_{t_f}\\
           \gamma'_0= \gamma_{t_f}\\
           \delta'_0= \delta_{0}
    \end{aligned}  
\end{equation}
where $t_f$ is the final time step of the symmetry-related segment in the first pulse half.
An important consideration with the real-time decompression of spline data is to guarantee that the signal is continuous between segments (something we refer to as "stitching"). In other words, the sampling rate must be maintained. In a pipelined implementation this means the data for the new segment should be supplied in advance dependent on the pipeline stage at which the data is used in the computation. The presented symmetric algorithm requires extra stitching between the two parts of the pulse, with the difference that the coefficients used in the second part are generated at the last iteration instead of being pre-stored in memory.
The timing diagram in Figure \ref{fig:timing_symmetric_pulses} shows that the coefficients used in the backward recursion are not all readily available at the same clock cycle and must be gradually introduced into the pipeline upon instruction from the control logic. The diagram shows that it is possible to update the coefficients in time to avoid any gap in the sampling rate of the output signal.

\subsection{Examples}
\label{sec:examples}
\subsubsection{Fitting errors for different pulse shapes}
\begin{figure}[ht!]\includegraphics[width=0.99\linewidth]{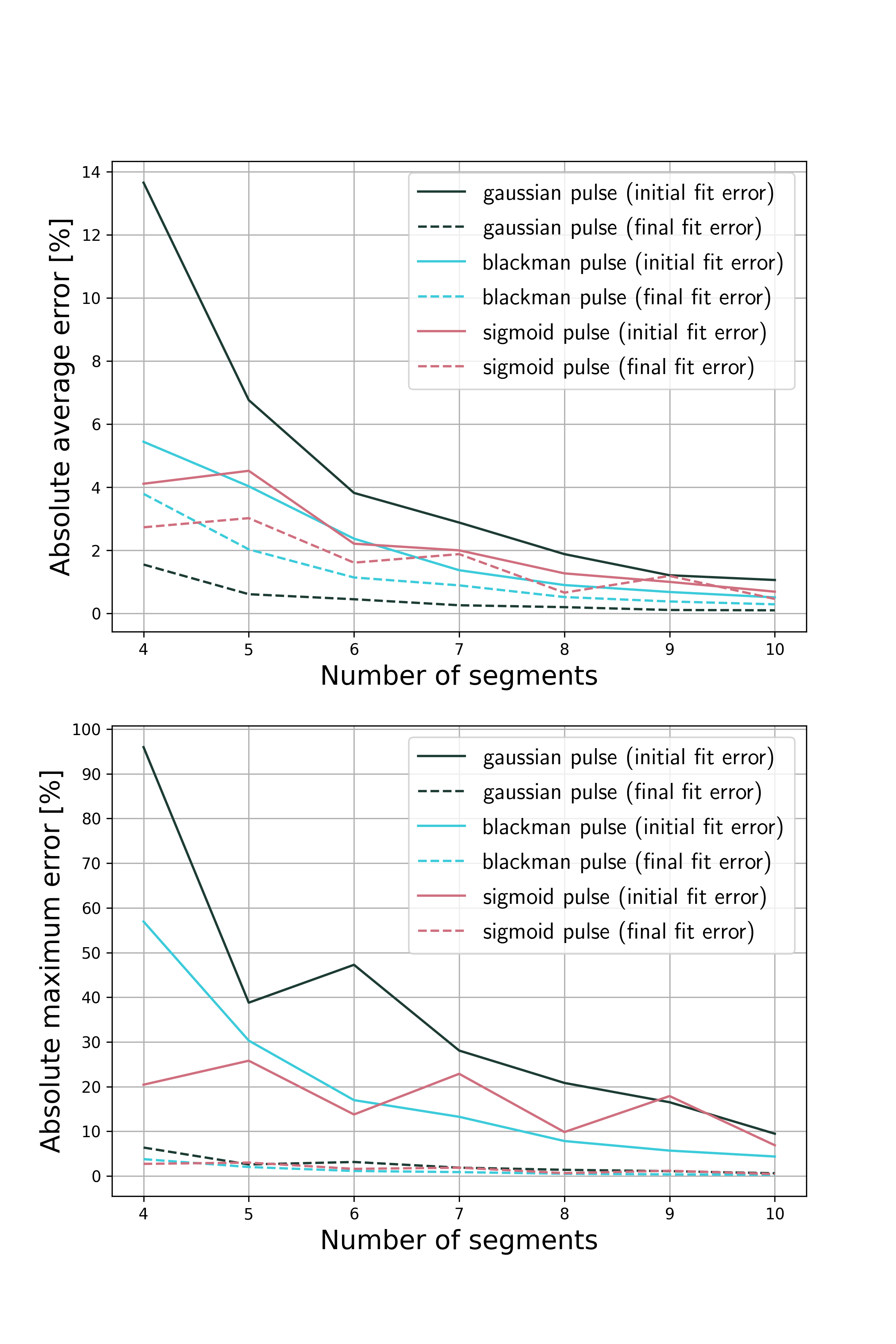}
\caption{Average and Maximum fitting errors for Gaussian, Blackman, and sigmoid functions. We compare the average and maximum fitting errors before and after applying the quantisation-aware optimisation on different functions, depending on the number of segments used for the piecewise cubic polynomial fit. One observes a general decrease in errors when increasing the number of segments. The maximum error is greatly reduced for all cases after applying the optimised fitting routine even when the original error is large and even for large segments indicating its merit. All pulses have 40000 samples. \label{fig:example_errors}}
\end{figure}

See Figure \ref{fig:example_errors}.

\subsubsection{Example of Bowler coefficients before and after optimisation}
\begin{table}[h]
\begin{tabular}{l|l|l}
     & Initial Coefficients (5 s.d. ) & \multicolumn{1}{l|}{Optimised Coefficients (5 s.d.)} \\ \hline
$\alpha_{s2}$ & $8821.7$ & $8821.7$  \\ \cline{1-1}
$\beta_{s2}$ & $3.7681$  & $ 3.\mathbf{3365}$ \\ \cline{1-1}
$\gamma_{s2}$ & $0.0012048$ & $0.00\mathbf{18556}$ \\ \cline{1-1}
$\delta_{s2}$ & $-6.2639e-07$ & $-6.\mathbf{3287}e-07$ \\ \cline{1-1}
$\alpha_{s3}$ & $27812$ & $27812$ \\ \cline{1-1}
$\beta_{s3}$ &  $3.1786$ & $3.1\mathbf{805}$ \\ \cline{1-1}
$\gamma_{s3}$ & $-0.0014799$ & $-0.0014\mathbf{820}$ \\ \cline{1-1}
$\delta_{s3}$ & $-9.3881e-10$ & $\mathbf{2.7533e-07}$\\ \cline{1-1}
\end{tabular}
\caption{Example of Bowler coefficients for initial and optimised fits in Fig. \ref{fig:optimised_fit}. The Bowler coefficients are defined in Eq. \ref{eq:bowler} and the subscript denotes the segment index. It is evident that the modifications of the linear, quadratic, and cubic polynomial coefficients are not of the order of several orders of magnitude, but were modified just enough to compensate for the cumulative errors. Here we have chosen the third and fourth segments for reference. The Gaussian, which is partitioned into 7 segments, is defined over a range of 20 $\mu$s and reads $y(t)=\sin(2 \pi ft)e^{-(t-10000)^2/(4\times10^6)}$. }
\end{table}

\end{document}